\documentclass{ifacconf}

\usepackage{graphicx}      % include this line if your document contains figures
\usepackage{natbib}        % required for bibliography
\usepackage{blindtext}
\usepackage{mathtools}
\usepackage{amsmath}
\usepackage{units}
\usepackage{url}
\usepackage{amssymb}
\usepackage{amsfonts}
\usepackage{mathrsfs}
\usepackage{csquotes}
\usepackage{subfig}
\usepackage{color}
\usepackage{tabularx} 
\usepackage{supertabular}
\usepackage{array,amsmath}
\usepackage[utf8x]{inputenc}

\usepackage{array,tabularx}
\usepackage{caption}   % for \captionof
\usepackage{textcomp}  % 
\usepackage[gen]{eurosym}
\newtheorem{problem}{Problem}

\newcolumntype{L}[1]{>{\raggedright\arraybackslash}p{#1}}
\newcolumntype{Y}{>{\arraybackslash}X}

\newif\ifmargincomments %A quick way of turning off margin comments for, say, arXiv submission
\margincommentstrue
\ifmargincomments

\else

\fi

\begin{document}
\begin{frontmatter}

\title{Optimal Sizing of Charging Energy Hubs for Heavy-Duty Electric Transport through Co-Optimization\thanksref{footnoteinfo}} 

\thanks[footnoteinfo]{This work was supported by RVO within the Charging Energy Hubs project with number
NGFS23020 and was partially funded by the Dutch National
Growth Fund.}

\author[First]{M. Izadi, D. Fernandez Zapico, M. Salazar, T. Hofman}

\address[First]{Mechanical Engineering Department, Eindhoven University of Technology, 
   the Netherlands (e-mail:\{m.izadi.najafabadi1, d.fernandez.zapico,  m.r.u.salazar, t.hofman\}@tue.nl).}

\begin{abstract}                % Abstract of 50--100 words
Electrification of heavy-duty vehicles places substantial stress on distribution grids, and Charging Energy Hubs (CEHs) mitigate these impacts by integrating charging infrastructure with renewable energy sources and battery storage. Optimal sizing of CEH components is therefore a critical investment decision, yet challenging because design choices depend strongly on operational dynamics. This work presents a mixed-integer linear programming model for the optimal sizing of CEH components, using a co-design approach that jointly optimizes component sizing and operational decisions. A case study for a heavy-duty fleet demonstrates the effectiveness of the method for cost-efficient, scalable, and grid-compliant CEH planning.
\end{abstract}


\end{frontmatter}
%===============================================================================

\section{Introduction}
Driven by regulations such as the EU Fit for 55 (\cite{EUFitFor55}), the Net Zero by 2050 targets (\cite{IEANetZero2050}), and
the rollout of Zero-Emission Zones in Dutch cities (\cite{ZEZDutchGov}), alongside the Netherlands’ national
goal to achieve climate neutrality by 2050, the logistics sector is being pushed to
transition toward electrification. This transition is expected to significantly increase
electricity demand, particularly for heavy-duty electric vehicles, due to larger
battery capacities and the need for rapid on-route charging. The shift is especially critical
for heavy-duty transport, as the Total Cost of Ownership (TCO) of electric trucks is
expected to be lower than that of diesel trucks by 2030. However, the Netherlands is experiencing grid congestion due to the rapid growth of both renewable energy
and electric vehicle charging demand, which, without solutions, could delay or even block
the transition to zero-emission heavy transport (\cite{TNO2025QueueGrid}).

\begin{figure}
\begin{center}
\includegraphics[width=6cm]{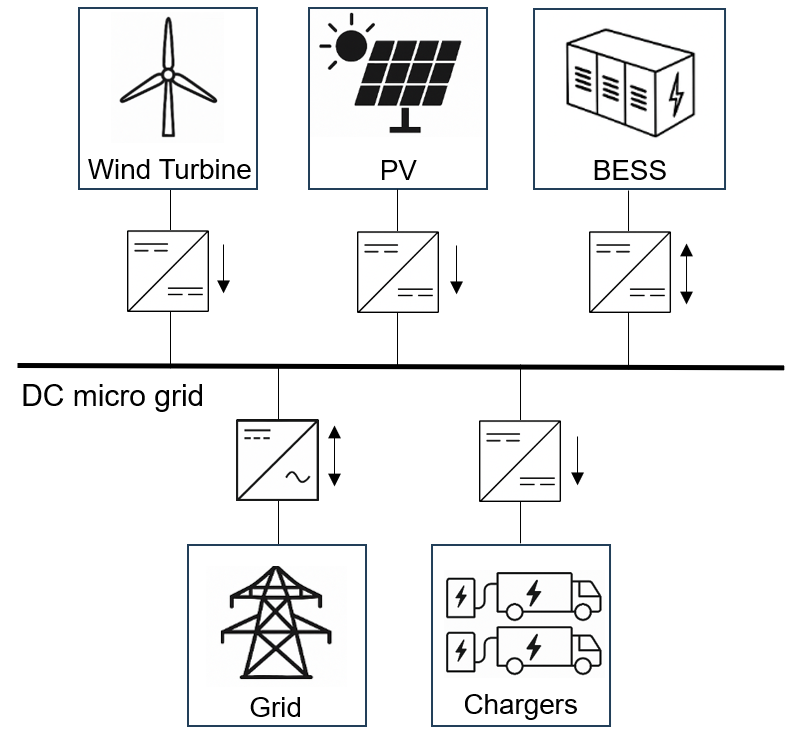}    % The printed column width is 8.4 cm.
\caption{Schematic configuration of
the DC microgrid within the CEH, integrating PV systems, WTs, BESS, and EV chargers through AC/DC and DC/DC converters.
} 
\label{fig:config}
\end{center}
\end{figure}

To address the challenges of rising electricity demand and increasing grid congestion,
the concept of Charging Energy Hubs (CEHs) has emerged as a promising solution.
CEHs are local energy systems that integrate charging infrastructure, battery storage,
and renewable energy sources within the existing grid, enabling smarter and more
flexible energy management. 
A direct current (DC) microgrid within a CEH facilitates the integration of local power sources, such as photovoltaic (PV) systems, wind turbines (WTs), and battery energy storage systems (BESS), with electric vehicle (EV) charging infrastructure through a unified DC distribution network. Native DC technologies such as PV systems and EV batteries make DC architectures attractive by reducing conversion losses, simplifying grid interfacing, and providing higher efficiency, robustness, reliability, and simpler control than AC-based systems. The configuration of the DC microgrid within the CEH is illustrated in Fig.\ref{fig:config}. It includes bidirectional AC/DC converters for grid interconnection, unidirectional DC/DC converters for interfacing the PV system, WTs, and chargers, and bidirectional DC/DC converters for the BESS to enable both charging and discharging operations.
CEHs can reduce grid power peaks by
increasing self-consumption of locally generated energy and, through their connection to
the national grid, provide flexibility during surpluses or shortages of renewable power.
Therefore, they play a key role in enabling the next phase of electrification in a
sustainable and scalable manner. 
A key aspect of CEH investment decisions is the sizing problem: determining the optimal number of chargers, battery storage units, and renewable energy sources to minimize TCO, while assuming a fixed battery capacity for each EV type. This objective can be extended to incorporate additional targets, e.g., emissions reduction and quality of service metrics for CEH users, such as waiting time.

\textit{Related literature:} Methods used in the literature for sizing charging infrastructure can be grouped into Physical–Economic Models (PEM) and data-driven Models. These categories refer to the underlying modeling approach used to estimate system behavior and evaluate different design options. Data-driven methods (\cite{WeiHanEtAl2022}) have gained attention for their ability to learn system behavior from historical data, but they generally do not guarantee global optimality, may require extensive training time, and often depend on the availability of sufficiently rich training datasets.
PEM approaches include rule-based and optimization-based methods. Rule-based strategies (\cite{MohandesAcharyaEtAl2020}) are structured and transparent; however, they generally fail to achieve optimal solutions and rely on predefined decision rules to define sizing strategies. Optimization-based PEM approaches include global programming methods and heuristic algorithms, such as genetic algorithms (\cite{TranHegazyEtAl2018}) and particle swarm optimization (\cite{BhattiSalamEtAl2019}). Although widely used, heuristic algorithms cannot guarantee global optimality and often require high computational effort.
Among global programming methods, linear programming and mixed-integer linear programming (MILP) are particularly attractive because they can guarantee global optimality and are supported by mature solvers with high computational efficiency. MILP has therefore become a dominant framework in large-scale energy system planning, as most system characteristics can be represented with reasonable accuracy using linear formulations.
In \cite{KhaksariTsaousoglouEtAl2021}, a MILP model is used to size an EV charging station by determining the optimal number and types of chargers while minimizing infrastructure cost under a quality of service requirement. The quality of service is enforced through chance constraints that limit the probability of charging delays beyond the requested departure time. However, the study does not consider the sizing of renewable generation or stationary battery storage and focuses solely on charger sizing.
In \cite{AluisioDicoratoEtAl2019}, a MILP framework is developed to size key microgrid components including PV, BESS, grid connections, and converters, but the number of chargers is determined separately through a rule-based approximation based on peak vehicle presence. %To manage computational complexity over a one-year operational horizon, the study clusters the days of the year into a limited set of representative scenarios that capture seasonal variability in solar generation and EV charging demand.
In \cite{Aliamooei-LakehCorzineEtAl2024}, a MILP model is proposed for sizing and siting grid-connected charging infrastructure in a distribution system, optimizing PV, WT, BESS, and EV charging stations. However, component sizes are modeled as continuous variables rather than discrete, which improves computational efficiency but limits physical scalability and modularity.
\cite{VehlhaberSalazarEtAl2024} propose a MILP framework for sizing grid-independent electric airports within a network intended for electric aviation. However, their model does not size the charging infrastructure and treats the PV and BESS capacities as continuous variables, thereby neglecting modular configuration. In conclusion, to the best of the authors’ knowledge, no existing study offers a MILP framework that discretely sizes all key CEH components and jointly optimizes design and operation, requiring the modelling of all component behaviors as well as every individual charging session, while also accounting for grid constraints.

\textit{Statement of contributions:} To address this gap in the literature, this work proposes a MILP model for the component sizing of CEHs to minimize the TCO for heavy-duty electric vehicles while ensuring that the EV fleet fully satisfies the predefined logistics demand in the studied zones. We adopt a co-design approach, in which component sizing and operational decisions are optimized jointly to capture the strong coupling between design choices and CEH operation. The model adopts a modular and scalable design approach to determine the optimal size of PV, WT, BESS, and chargers while accounting for power demand, weather profiles, energy prices, and grid constraints. All individual EV charging sessions are modeled using a smart-charging framework in which each session must occur within the vehicle’s arrival–departure window. The proposed MILP model incorporates the grid-power constraint, which imposes upper bounds on electricity import and export as defined by the CEH’s grid-connection contract.
\section{Methodology and Mathematical Model}
This study formulates the CEH design optimization problem as a MILP model to determine the number and types of charging units, as well as the optimal numbers of PV systems, WTs, and BESS units. A co-design framework is used in which the plant design (component sizing) and the energy management strategy, which distributes power among the different energy sources, are optimized simultaneously. This approach is important because sizing decisions strongly influence operational flexibility and costs, while operational constraints such as grid limits, renewable variability, and EV logistics directly determine which component sizes are feasible or cost-optimal. The operational variables include grid power exchange, battery charging and discharging power, battery state of charge (SOC), chargers power, and the decisions associated with charging/discharging the battery and importing/exporting power from/to the grid, as well as the EV–charger assignment decisions, which specify the allocation of each EV to a charger over time. The objective function minimizes TCO of the CEH while ensuring that the energy demand of the EV fleet is fully satisfied.
We adopt a modular configuration to ensure practical implementability and scalability. In this configuration, each component is treated as a discrete unit with known capacity and technical characteristics, and the optimization then selects the number of units to install.

Formulating the optimization problem over a one-year horizon results in a high-dimensional and computationally demanding MILP model. To reduce the computational burden, the methodology employs a scenario-based approach. We define a set of representative scenarios $s \in S$ because solar and wind power generation depend on weather conditions and seasons, and EV charging demand varies between weekdays and weekends as well as across different periods of the year. Each scenario represents a typical day and is assigned a corresponding occurrence factor $D_s$, indicating the number of times it occurs within the year. 
A scenario $s$ spans a full representative day, discretized into time slots $t \in T$ with resolution $\Delta t$, such that $|T| = 24 / \Delta t$. Operational variables use $(t,s)$ as indices, whereas design variables remain scenario-independent.

The model requires input data over a one-year planning horizon, including: (i) logistics data such as EV arrival and departure times and and energy demand of the EV fleet; (ii) weather data comprising solar radiation and wind speed profiles at the CEH location; (iii) grid constraints defining the maximum import and export power; (iv) electricity price data; and (v) technical and economic parameters describing the specifications and costs of the system components.
The model outputs include both optimal design and operation variables. The discrete design variables correspond to the number of PV units, WTs, BESS units, and charging points of each charger type, while the operational outputs include the grid power exchange, aggregate CEH demand, BESS charge and discharge power, battery SOC, and the optimal charging schedule. 

The following subsections detail the mathematical formulation of all CEH components, including their operational and physical constraints and the objective function. All models are expressed in linear form to ensure the tractability of the MILP model.
The CEH model uses the index sets $p \in P$ for PV, $w \in W$ for WT, and $b \in B$ for BESS to represent the available technology options..
 %Section \ref{sec:Nomenclature} presents all model parameters and decision variables in detail.
\subsection{PV Power Generation}
The total power generated by all PV technologies at time $t$ in scenario $s$, denoted by $P^{\mathrm{PV}}_{t,s}\,[\mathrm{kW}]$, is given by
\begin{equation}
P^{\mathrm{PV}}_{t,s}
   = n_p \sum_{p \in P} \eta_p A_p G_{t,s},
\label{pv model}
\end{equation}
where $n_p$ is the number of units of PV technology $p$, 
$G_{t,s}$ is the solar irradiance at time $t$ in scenario $s$ [$\mathrm{kW/m^2}$], 
and $\eta_p$ and $A_p$ denote the efficiency and area [$\mathrm{m^2}$] of technology $p$, respectively.
The effects of panel tilt angle, orientation, and ambient temperature on PV output are neglected. We assume that \(G_{t,s}\) represents the effective incident radiation and treat the PV efficiency \(\eta_p\) as constant.

 \iffalse
 over time.
 The PV installation is constrained by the available roof or land area as
\begin{equation}
   \sum_{p \in P} n_p A_p  \le S^{\mathrm{PV}}.
\end{equation}
\fi

\subsection{Wind Power Generation}
The total wind power generated at time $t$ in scenario $s$, denoted by
$P^{\mathrm{WT}}_{t,s}\,[\mathrm{kW}]$, is given by
\begin{align}
P^{\mathrm{WT}}_{t,s} =
\begin{cases}
\displaystyle n_w\sum_{w \in W} 0.5 c_w \rho A^{\mathrm{sw}}_w v_{t,s}^3,& v^{\mathrm{ci}}_w \le v_{t,s} \le v^{\mathrm{r}}_w. \\[6pt]
\displaystyle n_w \sum_{w \in W} 0.5 c_w \rho A^{\mathrm{sw}}_w (v^{\mathrm{r}}_w)^3,& v^{\mathrm{r}}_w \le v_{t,s} \le v^{\mathrm{co}}_w, \\[6pt]
0,& \text{otherwise.}
\end{cases}
\end{align}
with $n_w$ is the number of units of WT technology $w$, 
$A^{\mathrm{sw}}_w$ is the swept rotor area of WT type $w$ [$\mathrm{m}^2$], 
and $v^{\mathrm{ci}}_w$, $v^{\mathrm{co}}_w$, and $v^{\mathrm{r}}_w$ 
denote the cut-in speed, cut-out speed, and rated speed of WT type $w$ [m/s], respectively.
$c_w$ represents the turbine efficiency coefficient, defined as $c_w = \frac{P^{\mathrm{r}}_w}{0.5\, \rho\, A^{\mathrm{sw}}_w (v^{\mathrm{r}}_w)^3}$, where $P^{\mathrm{r}}_w$ is the rated power of WT type $w$.
 The wind speed at the WT hub height, denoted $v_{t,s}$, is estimated from the measured wind speed $v^{\mathrm{m}}_{t,s}$ using the power-law relationship $v_{t,s} = v^{\mathrm{m}}_{t,s} \left(\frac{h_{\mathrm{hub}}}{h_{\mathrm{m}}}\right)^{\alpha}$, where $h_{\mathrm{m}}$ is the wind measurement height, and $\alpha$ is the wind shear exponent, typically around $0.143$ under neutral atmospheric stability.
 We apply this scaling because wind speed increases with height, so the measured data must be adjusted to represent conditions at the WT hub level. We assume a constant air density, thereby neglecting temperature and pressure effects.

\subsection{Battery Energy Storage System}
We represent the BESS dynamics through the evolution of its SOC over each time interval $\Delta t$, accounting for both charging and discharging processes. The SOC of the $b$-th BESS technology at time $t$ in scenario $s$ is given by
\begin{equation}
    E_{b,t,s} = E_{b,t-1,s}
    + \eta^{\mathrm{ch}}_{b} P^{\mathrm{ch}}_{b,t,s}\Delta t
    - \frac{P^{\mathrm{dis}}_{b,t,s}}{\eta^{\mathrm{dis}}_{b}}\Delta t
    - z_b n_b s_b \Delta t,
\end{equation}
where $s_b$ is the unit size of BESS type $b$ [kWh], $\eta^{\mathrm{ch}}_{b}$ and $\eta^{\mathrm{dis}}_{b}$ are its charging and discharging efficiencies, %$P^{\mathrm{ch}}_{b,t,s}$ and $P^{\mathrm{d}}_{b,t,s}$ denote its charge and discharge power [kW],
and $z_b$ is its self-discharge rate [$\mathrm{h}^{-1}$]. Two non-negative variables, $P^{\mathrm{ch}}_{b,t,s}$ and $P^{\mathrm{dis}}_{b,t,s}$, are introduced to represent charging and discharging power separately. This formulation allows us to apply different charging and discharging efficiencies and preserves linearity in the SOC equation.
The model enforces the BESS operational limits through the constraint
\begin{equation}
n_b s_b\,\underline{E}_b \le E_{b,t,s} \le n_b s_b\,\overline{E}_b,
\end{equation}
with \(\underline{E}_b\) and \(\overline{E}_b\) denoting the minimum and maximum allowable SOC fractions, respectively, satisfying
\(0 \le \underline{E}_b \le \overline{E}_b \le 1\).
To ensure consistency across representative days, we constrain the battery SOC at the beginning of each day to start from the same initial value:
\begin{equation}
     E_{b,1,s}=n_b s_b E_{\mathrm{o}},
\end{equation}
where $n_b$ is the number of units of BESS technology $b$ and $E_{\mathrm{o}}$ is initial SOC fractions within $[0~1]$. The charging and discharging power of the battery must be non-negative and bounded by the maximum charge and discharge limits, $P_{b}^{\mathrm{max,c}}$ and $P_{b}^{\mathrm{max,d}}$, respectively:
\begin{equation}
   0 \le P^{\mathrm{ch}}_{b,t,s} \le n_b P_{b}^{\mathrm{max,c}},
\end{equation}
\begin{equation}
   0 \le P^{\mathrm{dis}}_{b,t,s} \le n_b P_{b}^{\mathrm{max,d}}.
\end{equation}
To avoid simultaneous charging and discharging within the same time step, we impose the following constraints:
\begin{equation}
   0 \leq P^{\mathrm{ch}}_{b,t,s} \leq N_b^{\mathrm{max}} P_b^{\mathrm{max,c}} \, \delta_{t,s},
\end{equation}
\begin{equation}
   0 \leq P^{\mathrm{dis}}_{b,t,s} \leq N_b^{\mathrm{max}} P_b^{\mathrm{max,d}} \, (1 - \delta_{t,s}),
\end{equation}
where $\delta_{t,s}$ is a binary variable indicating whether the BESS is charging (1) or discharging (0) at time $t$ in scenario $s$, and $N_b^{\mathrm{max}}$ denotes the maximum number of units of BESS type $b$ that can be installed within the CEH.

\subsection{Smart Scheduling Model for EV Charging Sessions}
This work considers different types of chargers, characterized by their rated power, as potential candidates for installation in CEH. A set of candidate chargers is defined as $c \in C$, and for each charger in this set, a binary decision variable $q_c$ is introduced to indicate whether the charger is selected for installation ($q_c = 1$) or not ($q_c = 0$). Considering $N_c$ types of chargers, let $m_l$ denote the number of chargers of type $l$ in the set of candidate chargers for all $l \in \{1, 2, \ldots, N_c\}$. One reasonable assumption could be to set $m_l$ equal to the average number of vehicles parked simultaneously in the CEH over the planning horizon.

We consider a set of arriving vehicles $v \in V$, where each vehicle $v$ is characterized by its arrival time $t^{\mathrm{arr}}_{v}$, departure time $t^{\mathrm{dep}}_{v}$, and charging energy requirement $E_v$. All of these logistic charging data are provided as inputs to the model. To model EV charging sessions in the CEH, the approach is inspired by \cite{KhaksariTsaousoglouEtAl2021}, employing a smart scheduling model in which each EV can start charging at a time slot after its arrival and complete its charging session before departure. The charging process must be continuous; once it starts, it cannot be interrupted.

We index candidate start times of charging sessions using a relative time 
variable $t_r \in T_r$, where $T_r := \{1,\ldots, T_{\max}^{\mathrm{parked}}\},~ T_{\max}^{\mathrm{parked}} :=\max\limits_{v}(t^{\mathrm{dep}}_{v}-t^{\mathrm{arr}}_{v}),$ with $T_{\max}^{\mathrm{parked}}$ denoting the maximum parking duration 
observed in the logistics data.
Binary variables $x_{v,c,t_r} \in \{0,1\}$ indicate whether vehicle $v$ starts charging at charger $c$ at relative
index \(t_r\) (\(x_{v,c,t_r}=1\)) or not (\(x_{v,c,t_r}=0\)); equivalently, \(x_{v,c,t_r}=1\) means $v$ starts charging at charger $c$ at time slot $t^{\mathrm{arr}}_{v}+t_r-1$. The relative time variable $t_r$ is introduced to reduce the problem size by defining $x_{v,c,t}$ only over a limited window after each vehicle’s arrival, thereby decreasing the number of binary variables while preserving charging–schedule accuracy.

The first constraint ensures that a charger $c$ from the set of candidate chargers can be assigned to a vehicle only if it is selected for installation:
\begin{equation}
   x_{v,c,t_r,s} \le q_c, \quad \forall\, c \in C,\, v \in V,\, t_r \in T_r.
\end{equation}
If $v$ is assigned to $c$, then $\tau_{v,c}$ time slots are required to complete its charging task, $\tau_{v,c} = \frac{1}{\Delta t} 
    \left\lceil 
    \frac{E_v}{p_{v,c}} 
    \right\rceil,$ where $\lceil \cdot \rceil$ denotes rounding up to the nearest integer, and $p_{v,c} = \min\{p^{\max}_v,\, p^{\max}_c\}$ is the effective charging rate, with 
$p^{\max}_v$ being the maximum charging rate of vehicle $v$. For simplicity, the model assumes a constant charging power during each charging session.

For each scenario 
$s$, each EV is assigned to exactly one charger and starts charging at exactly one time slot $t \in T$:
\begin{equation}
    \sum_{c \in C} \sum_{t_r = 1}^{t^{\mathrm{dep}}_v - t^{\mathrm{arr}}_v - \tau_{v,c} + 2} 
    x_{v,c,t_r,s} = 1, 
    \quad \forall v \in V.
\end{equation}
Furthermore, for all chargers $c\in C$ , time slots $t\in T$, and scenarios $s\in S$, at most one EV can be charging on a charger at any given time:
\begin{equation}
  \sum_{v \in V} ~~
  \sum_{t_r = \max\{1,\, t - t^{\mathrm{arr}}_v - \tau_{v,c} + 2\}}^{t - t^{\mathrm{arr}}_v + 1}
  x_{v,c,t_r,s} \le 1.
\end{equation}
Therefore, for all chargers $c \in C$ and time slots $t \in T$ in scenario $s$, the charging power of charger $c$ is expressed as:
\begin{equation}
    P_{c,t,s} = 
    \sum_{v \in V} 
    p_{v,c}
    \sum_{t_r = \max\{1,\, t - t^{\mathrm{arr}}_v - \tau_{v,c} + 2\}}^{t - t^{\mathrm{arr}}_v + 1} 
    x_{v,c,t_r,s}.
\end{equation}
The set of chargers $C$ includes all candidate chargers of different types. 
For each charger type $l \in \{1, \ldots, N_c\}$, a subset $C_l \subset C$ is defined and ordered as $C_l = \{c_l(1), \ldots, c_l(m_l)\}$. 
To avoid symmetric solutions that differ only by a permutation of identical chargers, 
the following symmetry-breaking constraints are imposed for all 
$l \in \{1,\ldots,N_c\}$ and $j \in \{1,\ldots,m_l-1\}$:
\begin{equation}
q_{c_l(j+1)} \leq q_{c_l(j)}, 
\end{equation}

\begin{equation}
x_{v,\,c_l(j+1),\,t,s} \leq 
\sum_{t_r = \max\{t_v^{\mathrm{arr}},\, t - \tau_{v,\,c_l(j)} + 1\}}^{t} 
x_{v,\,c_l(j),\,t_r,s}, 
\forall v \in V.
\end{equation}
The first constraint enforces a sequential installation order within each charger type: charger $j+1$ cannot be installed unless charger $j$ is also installed. The second ensures that a higher-index charger of the same type can only be used once all lower-index chargers are already in use.
\subsection{Grid Constraints and Power Balance }
\vspace{-0.2 cm}
The grid power constraint imposes time-dependent bounds on electricity import and export according to the CEH’s grid connection agreement. The grid power at time $t$ in scenario $s$, denoted by $P^{g}_{t,s}\,[\mathrm{kW}]$, must satisfy
\begin{equation}
    -\bar{P}^{g,\mathrm{inj}}_{t,s} \le P^{g}_{t,s} \le \bar{P}^{g,\mathrm{wdl}}_{t,s},
\end{equation}
 where $\bar{P}^{g,\mathrm{inj}}_{t,s}$ and $\bar{P}^{g,\mathrm{wdl}}_{t,s}$ define the time-varying injection and withdrawal limits [kW], which may vary with operational conditions such as day--night or weekday--weekend periods. Positive values of $P^{g}_{t,s}$ represent grid withdrawal, while negative values represent grid injection.

The grid trading cost, $C^{\mathrm{el}}_{t,s}$, [\euro{}], represents the monetary exchange associated with 
electricity trading between the CEH and the grid. It is positive when electricity is purchased 
from the grid 
($C^{\mathrm{el}}_{t,s} = \lambda^{\mathrm{b}}_{t,s} P^{g}_{t,s} \Delta t$) 
and negative when electricity is sold to the grid 
($C^{\mathrm{el}}_{t,s} = \lambda^{\mathrm{s}}_{t,s} P^{g}_{t,s} \Delta t$). 
Here, $\lambda^{\mathrm{b}}_{t,s}$ and $\lambda^{\mathrm{s}}_{t,s}$ denote the 
purchase and selling prices of electricity at time $t$ in scenario $s$ [\euro{}/kWh].

Assuming $\lambda^{\mathrm{b}}_{t,s} > \lambda^{\mathrm{s}}_{t,s}$, we relax these two equality constraints to the following inequalities:
\begin{align}
 C^{\mathrm{el}}_{t,s} &\geq \Delta t\lambda^{\mathrm{b}}_{t,s} P^{\mathrm{g}}_{t,s},\\
 C^{\mathrm{el}}_{t,s} &\geq \Delta t\lambda^{\mathrm{s}}_{t,s} P^{\mathrm{g}}_{t,s}.
\end{align}
This relaxation avoids the need for binary variables to distinguish import and export modes. Since $C^{\mathrm{el}}_t$ is penalized in the objective function, the optimal solution at each time step occurs when one of these two constraints is active.

The power balance constraint ensures that, at each time step $t$ in scenario $s$, the total power generation and supply, including PV and wind generation, possible BESS discharge, and grid withdrawal, equals the total demand within the CEH, which consists of EV charging loads, possible BESS charging, and grid injection. The corresponding power balance equation is given by:
\begin{equation}
P^{\mathrm{pv}}_{t,s} + P^{\mathrm{WT}}_{t,s}+P^{\mathrm{g}}_t + \sum_{b \in B} P_{b,t,s}^{\mathrm{dis}}-\sum_{b \in B} P_{b,t,s}^{\mathrm{ch}}= \sum_{c\in C} P_{c,t,s}.
\label{balance cons}\end{equation}

\subsection{Objective Function}
The objective of the optimization problem is to minimize the TCO associated with the design and operation of the CEH over a one-year horizon, expressed as
\begin{equation}
  \min J^{\mathrm{TCO}} = J^{\mathrm{CapEx}} + J^{\mathrm{OpEx}} + J^{\mathrm{deg}}.
\end{equation}
The first term, $J^{\mathrm{CapEx}}$, represents the annual capital expenditures 
related to the investment costs of PVs, WTs, BESS units, and chargers. 
The second term, $J^{\mathrm{OpEx}}$, accounts for operational expenditures, including grid energy trading 
and maintenance costs. The third term, $J^{\mathrm{deg}}$, represents the BESS degradation cost, 
which penalizes excessive cycling of the BESS to profit from grid exports.

The capital expenditures are annualized using a capital recovery factor,
denoted by $\kappa$, which accounts for the time value of money and 
component lifetimes. This factor converts the investment cost of each 
component into an equivalent annual cost, based on the discount rate $r$ 
and its lifetime $L$, defined as
\begin{equation*}
    \kappa = \frac{r(1+r)^{L}}{(1+r)^{L} - 1}.
\end{equation*}
We compute a separate value of $\kappa$ for each technology type (PV, WT, BESS, and chargers) according to their lifetimes. The total capital expenditure is then given by 
\begin{align}
J^{\mathrm{CapEx}} =
&\sum_{p \in P} n_p\,\kappa_p C_p +
 \sum_{w \in W} n_w\,\kappa_w C_w + \\
&\sum_{b \in B} n_b\,\kappa_b C_b +
 \sum_{c \in C} q_c\,\kappa_c C_c ,
\end{align}
where $C_p$, $C_w$, $C_b$, and $C_c$ are the investment costs of a single unit 
of PV technology $p$, WT type $w$, BESS technology $b$, and charger type $c$ 
[\euro], respectively.
Each term represents the annualized investment cost of the corresponding CEH component. 

The operational expenditure aggregates the grid trading cost and the annual maintenance costs of the assets:
\begin{align}
    &J^{\mathrm{OpEx}} =J^{\mathrm{OpEx,G}}+J^{\mathrm{OpEx,M}}= \sum_{s\in S} D_s \sum_{t \in T} C^{\mathrm{el}}_{t,s}+ \nonumber \\
    & \sum_{p\in P} n_p C^{\mathrm{M}}_p  + \sum_{w\in W}  n_w C^{\mathrm{M}}_w   
    +\sum_{b\in B}  n_b C^{\mathrm{M}}_b   +\sum_{c\in C} q_c C^{\mathrm{M}}_c  , 
\end{align}
with the first term representing the annual grid trading cost as the sum of time-step grid exchange costs across all scenarios, weighted by their yearly occurrences $D_s$. The parameters $C^{\mathrm{M}}_p$, $C^{\mathrm{M}}_w$, 
$C^{\mathrm{M}}_b$, and $C^{\mathrm{M}}_c$ denote the annual maintenance cost per 
unit of PV technology $p$, WT type $w$, BESS technology $b$, and charger type $c$ 
[\euro], respectively.

We formulate the degradation cost of the BESS to penalize the battery power magnitude :
\begin{equation}
    J^{\mathrm{deg}}= \sum_{b \in B} \sum_{s \in S} D_s\sum_{t \in T} C_b^{\mathrm{deg}} ( P_{b,t,s}^{\mathrm{dis}}+P_{b,t,s}^{\mathrm{ch}}), 
\end{equation}
wher $C_b^{\mathrm{deg}}$ is degradation cost of BESS type $b$ [\euro/kW]. This term discourages excessive use of the BESS and mitigates over-cycling driven by grid-export incentives.

With all components, constraints, and the objective defined, the CEH design problem is formulated as follows.
\begin{problem}[CEH Design Optimization]
\label{prb:TCO}
\begin{align*}   
    \min_z~&~ \; J^{\mathrm{TCO}} = J^{\mathrm{CapEx}} + J^{\mathrm{OpEx}} + J^{\mathrm{deg}}\\
    \mathrm{s.t.}~&~\\
 &~(\ref{pv model})\text{--}(\ref{balance cons}),\\
 &~n_p,n_w,n_b \in \mathbb{N},\\
 &~q_c\in \{0,1\},~~~ \forall c \in C,\\
 &~x_{v,c,t_r,s}\in \{0,1\},~~~ \forall v\in V,~ c \in C,~ t_r\in T_r,~  s \in S,\\
 &~\delta_{t,s}\in \{0,1\},~~~ \forall  t\in T, ~ s \in S,
\end{align*} 
where $z = (z^{\mathrm{des}},z^{\mathrm{op}})$ collects all design and operational decision variables, with
\[
\begin{aligned}
 z^{\mathrm{des}}=& \{ n_p,\, n_w,\, n_b,q_c\},\\
 z^{\mathrm{op}} =& \{\, P^{\mathrm{ch}}_{b,t,s},\; P^{\mathrm{dis}}_{b,t,s},\; 
P_{c,t,s},\; P^{g}_{t,s},
 E_{b,t,s},\; \delta_{t,s},\; x_{v,c,t_r,s}, \,\\
 & \quad \forall c \in C, ~v\in V,~ t\in T ,~ t_r\in T_r,~ s \in S.\}
\end{aligned}
\]
\end{problem}
Problem~\ref{prb:TCO} is a MILP and therefore guarantees global optimality and is solvable with off-the-shelf solvers.

\section{Case Study}
\vspace{-0.2 cm}
In the case study, we use 12 monthly weather profiles together with 24 demand profiles (weekday and weekend for each month). Each scenario pairs the weather profile of a given month with its corresponding weekday or weekend demand pattern, yielding 24 representative scenarios that reflect typical CEH operating conditions throughout the year. Each scenario is assigned an occurrence factor $D_s$, corresponding to the number of times it occurs annually. All scenarios are simulated with a one-hour time resolution. Note that this scenario definition is based on the strong weekday/weekend pattern and monthly seasonality observed in the logistics charging data, as well as the monthly patterns present in the weather data.

The case study considers one technology per component: PV, WT, and BESS.
For the PV unit, the nominal power is $\unit[550]{W}$, with an efficiency of $\eta_p = 0.20$, a surface area of $A_p = \unit[2.58]{m^2}$, and 
the investment cost of $C_p = \euro{495}$.
The WT unit is characterized by a cut-in speed of $v^{\mathrm{ci}}_w = \unit[3]{m/s}$, 
a cut-out speed of $v^{\mathrm{co}}_w = \unit[20]{m/s}$, and a rated speed of $v^{\mathrm{r}}_w = \unit[13]{m/s}$. 
The swept area of the WT blades is $A^{\mathrm{sw}}_w = \unit[1734]{m^2}$, 
with an air density of $\rho = \unit[1.225]{kg/m^3}$ and a hub height of $h_w = \unit[60]{m}$. 
The rated power is $P^{\mathrm{r}}_w = \unit[500]{kW}$, and the investment cost is $C_w = \euro{750000}$ .
The BESS unit has a capacity of $s_b = \unit[580]{kWh}$, 
a maximum charge and discharge power of $P_b^{\mathrm{max,c}} = P_b^{\mathrm{max,d}} = \unit[300]{kW}$, 
and charge/discharge efficiencies of $\eta^{\mathrm{ch}} = \eta^{\mathrm{dis}} = 0.95$. 
The SOC is limited to $[\underline{E}_b, \overline{E}_b] = [0.1, 0.95]$, with an initial value of $E_{b,0} = 0.5$. 
The self-discharge rate is $z_b = \unit[10^{-4}]{{h}^{-1}}$, the investment cost is $C_b = \euro{32000}$, the degradation cost is $C_b^{\mathrm{deg}}=0.03$ \euro/kW, and the maximum number of installable units is $N_b^{\mathrm{max}} = 100$.
We consider two types of chargers in the study. 
For each type $l \in \{1, 2\}$, the number of candidate chargers is set to six, based on logistics data indicating that an average of six EVs are parked simultaneously in the CEH.
This results in a total of 12 candidate chargers (six per type). 
The maximum charging powers of the two charger types are 
$\unit[180]{kW}$, and $\unit[360]{kW}$, 
with respective investment costs of $\euro{90{,}000}$ and $\euro{180{,}000}$ per unit.
A single type of EV is considered in the analysis, with a maximum battery charging rate of $p^{\mathrm{max}}_v=\unit[400]{kW}$. The maximum allowable grid import and export power is set to 
$\unit[600]{kW}$ during daytime hours (08{:}00--20{:}00) 
and $\unit[800]{kW}$ during nighttime hours (20{:}00--08{:}00), 
reflecting grid connection limits for the CEH.
Component lifetimes are assumed to be 20~years for the PV and WT units, 
15~years for the BESS, and 10~years for the chargers. 
A discount rate of $r = 2.75\%$ is applied for the economic analysis. Annual maintenance costs for all components are modeled as $1\%$ of their investment cost. Note that the model parameters were selected in consultation with experts to 
reflect a realistic medium-size CEH and to remain consistent with typical 
values reported in manufacturers' datasheets.
The weather data and electricity price data for the year 2024 were obtained from \cite{weather} and \cite{entsoeAPI2024}, respectively. We used logistic charging data provided by Maxem, a Dutch company specializing in smart charging and energy-management solutions for EV infrastructure. The dataset consists of aggregated measurements collected directly from charging stations and associated metering equipment, processed through Maxem’s energy management platform.

We implement the proposed MILP model in \textsc{YALMIP} within the \textsc{MATLAB} environment, using the \textsc{Gurobi} optimizer (\cite{gurobi}). The simulation was performed on an HP ZBook workstation equipped with an Intel Core i7-13700H processor (2.40\,GHz) and 16\,GB of RAM.
Based on the MILP optimization, the resulting CEH configuration consists of 
\textbf{1110 PV units}, \textbf{1 wind turbine}, and \textbf{6 BESS units}. 
For the charging infrastructure, $q_c = [1,1,1,1,0,0,1,0,0,0,0,0]^\top$ indicates that \textbf{4 chargers of the first type} and \textbf{1 charger of the second type} are selected from the candidate set.
The optimized CEH exhibits a total annual cost of 
$\euro{4.01\times 10^{5}}$, of which $68.96\%$ corresponds to 
annualized capital expenditure, $17.87\%$ to grid trading costs, 
$8.49\%$ to annual maintenance, and $4.68\%$ to battery 
degradation.

\begin{figure}
\begin{center}
\includegraphics[width=9cm]{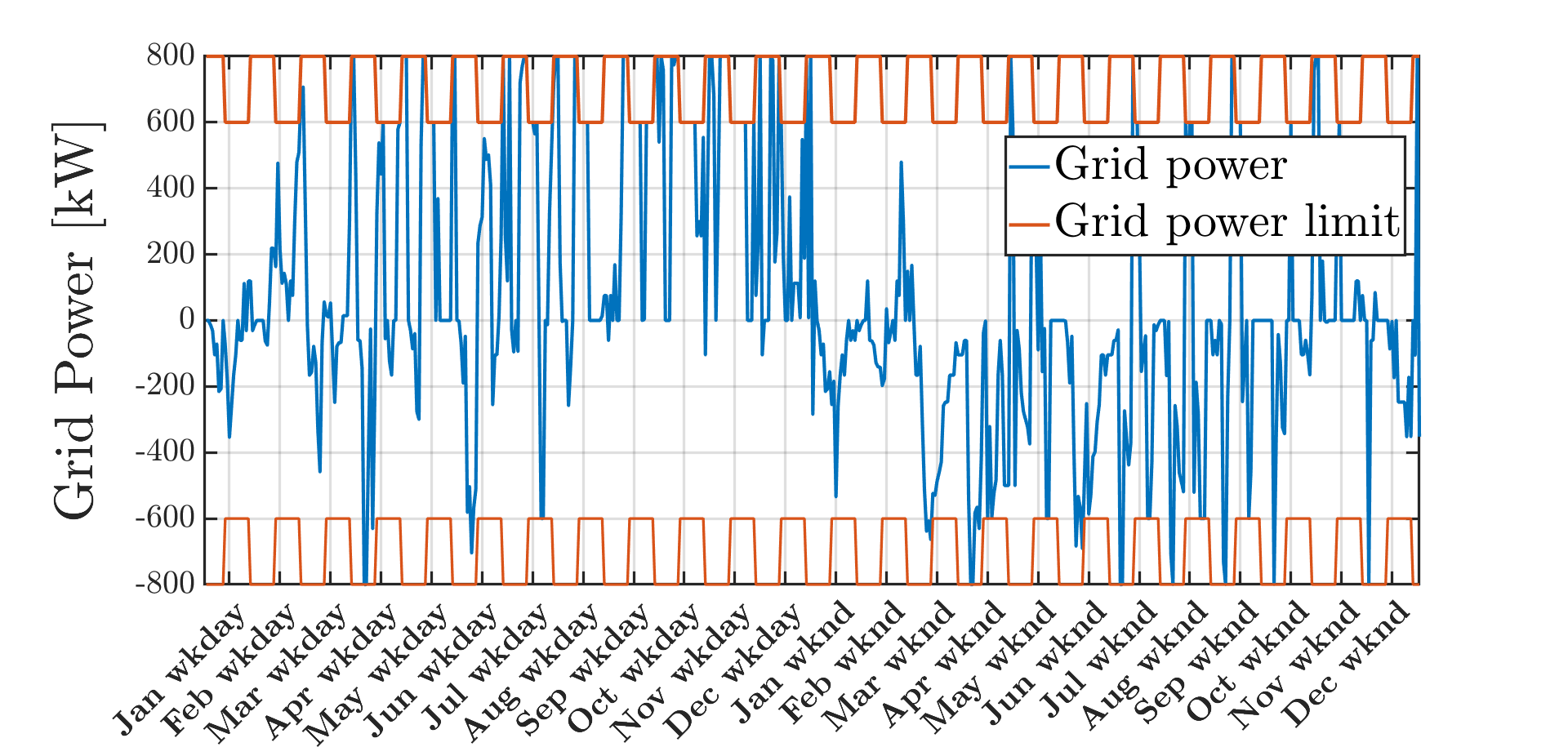}    % 
\caption{Grid power across all representative scenarios (positive: withdrawals, negative: injections) with grid limits shown in orange.%Grid power profile over all representative scenarios, with positive values indicating withdrawals and negative values indicating injections. Orange lines show the grid power limits.
} 

\label{fig:gridpower}
\end{center}
\end{figure}

\begin{figure}
\begin{center}
\includegraphics[width=9cm]{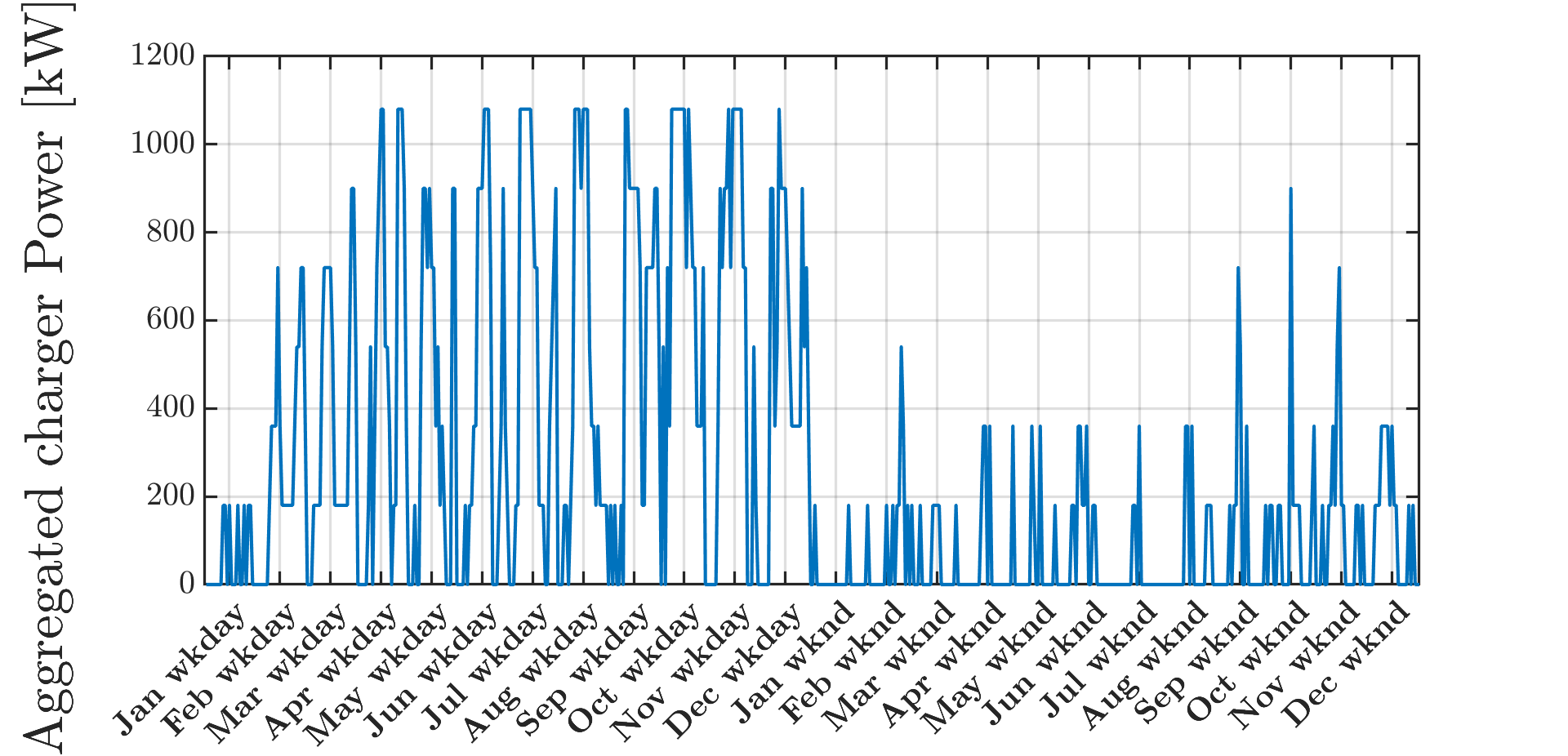}    % 
\caption{Aggregated EV charging power across all representative scenarios, showing charging-demand variability.%Aggregated EV charging power over all representative scenarios, illustrating the temporal variability of EV charging demand throughout the year.
} 
\label{fig:charger power}
\end{center}
\end{figure}

\begin{figure}
\begin{center}
\includegraphics[width=9cm]{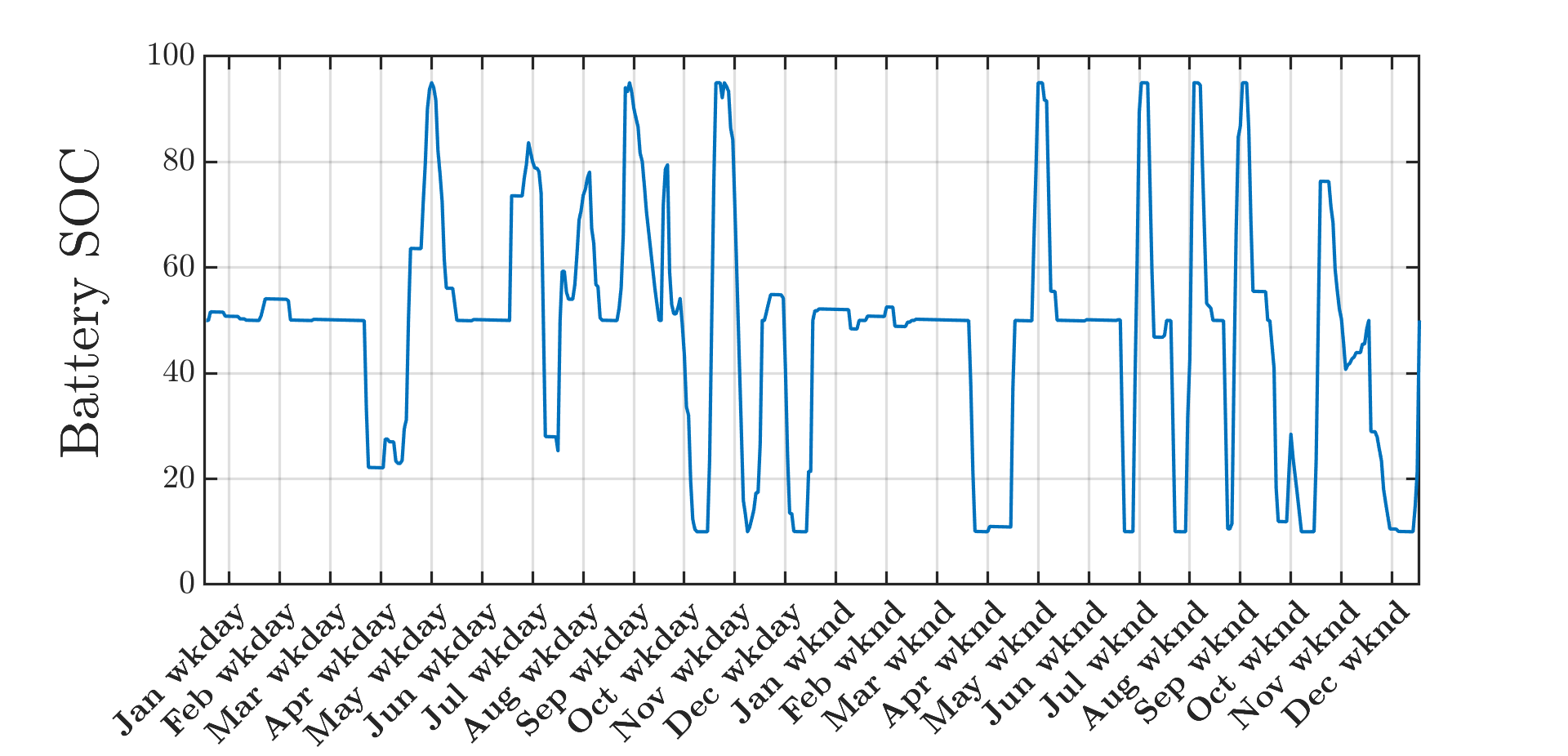}    % 
\caption{Battery SOC across all representative scenarios, each day starting from the same initial SOC
} 
\label{fig:soc}
\end{center}
\end{figure}
The optimal operational profiles for the CEH configuration are shown in Fig.~\ref{fig:gridpower}–\ref{fig:soc}. 
Fig.~\ref{fig:gridpower} illustrates the grid power exchange across all scenarios, showing that the import and export limits are strictly respected at all times. 
Fig.~\ref{fig:charger power} depicts the aggregated charging power delivered by the installed chargers, reflecting the EV demand dictated by the logistics data and the optimized charging schedule. 
Fig.~\ref{fig:soc} presents the BESS SOC trajectory; each representative day begins at a fixed SOC level of $50\%$, ensuring consistency across scenarios. 
Despite peak EV charging demands exceeding $\unit[1100]{kW}$, all logistics demands are met without violating the $\unit[600/800]{kW}$ daytime/nighttime grid limits, thanks to the coordinated use of renewable generation and battery storage.

\begin{figure}
\begin{center}
\includegraphics[width=9cm]{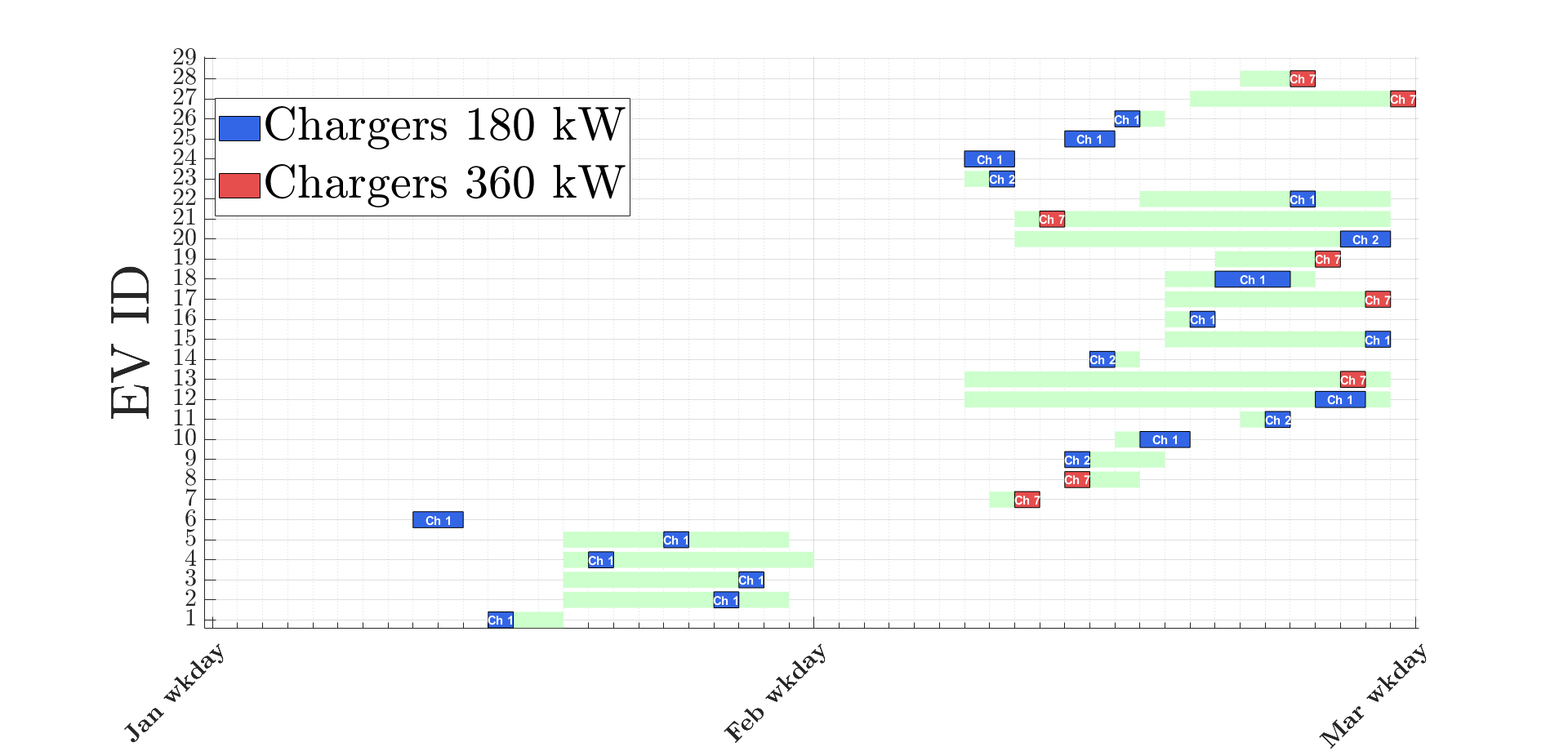}    % 
\caption{Zoomed-in view of the optimized EV charging schedule for the first two representative days.
} 
\label{fig:chargingplan}
\end{center}
\end{figure}

Fig. \ref{fig:chargingplan} illustrates the optimized charging schedule for the CEH.
The light-green bars represent each vehicle’s availability window, (arrival–departure interval). Each EV is assigned to exactly one charger, and the corresponding charging interval is shown as a colored block (blue: \unit[180]{kW}, red: \unit[360]{kW}) placed within its availability window. 
As the full-year schedule contains a large number of charging events across all representative days, the plot is zoomed in to display only the first two representative days for clarity.

\begin{figure}
\begin{center}
\includegraphics[width=9cm]{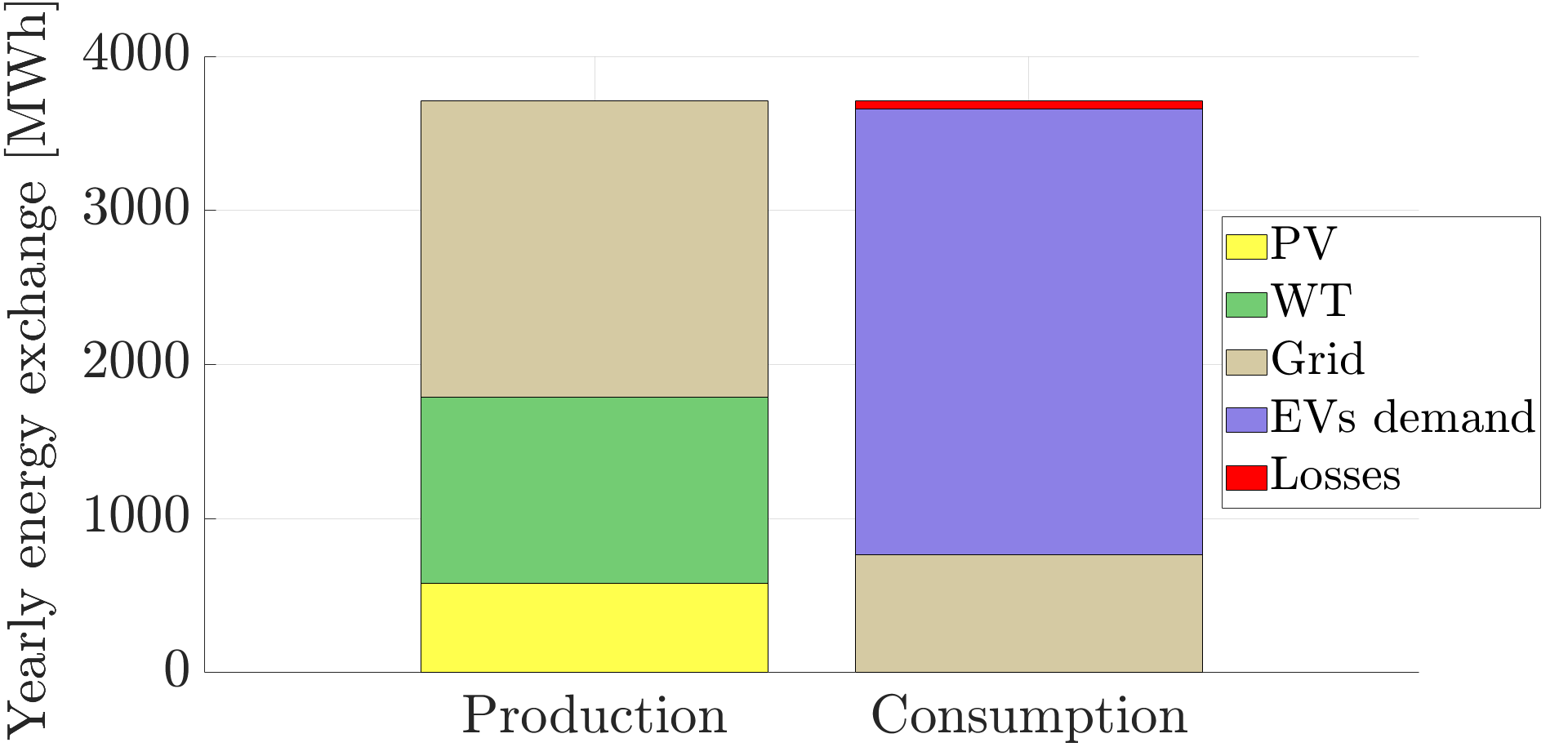}    % 
\caption{Breakdown of yearly energy production and consumption across all sources and uses.
} 
\label{fig:energyexchange}
\end{center}
\end{figure}
As shown in Fig.~\ref{fig:energyexchange}, the yearly energy balance indicates that PV generation contributes approximately $15.7\%$ of the total production, wind generation supplies around $32.5\%$, and the remaining $51.8\%$ is imported from the grid. On the consumption side, about $78\%$ of the energy is used to meet the EV demand, $20.7\%$ is exported back to the grid, and around $1.3\%$ corresponds to battery losses.
\section{Conclusions}
\vspace{-0.2cm}
This work proposed a mixed-integer linear programming (MILP) formulation for the cost-optimal sizing of Charging Energy Hub (CEH) components within a co-design framework, highlighting how sizing choices are fundamentally shaped by operational constraints such as grid limits and logistics-driven charging demand. By incorporating representative scenarios, the approach captures seasonal weather variability and charging patterns while keeping the problem computationally tractable. The case study demonstrates that a properly sized CEH can reliably serve all fleet charging needs without violating grid constraints, underscoring the potential of integrated renewable generation and storage to enable scalable, grid-compliant heavy-duty electrification.

Future work includes incorporating more realistic EV charging power profiles that decrease over the charging session, conducting sensitivity analyses to assess how key parameters affect optimal CEH sizing, and developing a CEH simulation tool for dynamic validation. Another promising direction is using the machine learning method from \cite{LiOuyangEtAl2023} to accelerate MILP solving by learning effective branch-and-cut separators.

\section*{DECLARATION OF GENERATIVE AI}
The authors used ChatGPT to improve the clarity and structure of several sentences in this work.

\bibliography{Izadi.Fernandez-Zapico.ea.IFAC26}             % bib file to produce the bibliography
%\bibliography{../../../bibliography/main,../../../bibliography/aviation,../../../bibliography/energyhubs,../../../bibliography/SML_papers}                                 % with bibtex (preferred)
                                                   
%\begin{thebibliography}{xx}  % you can also add the bibliography by hand

%\bibitem[Able(1956)]{Abl:56}
%B.C. Able.
%\newblock Nucleic acid content of microscope.
%\newblock \emph{Nature}, 135:\penalty0 7--9, 1956.

%\bibitem[Able et~al.(1954)Able, Tagg, and Rush]{AbTaRu:54}
%B.C. Able, R.A. Tagg, and M.~Rush.
%\newblock Enzyme-catalyzed cellular transanimations.
%\newblock In A.F. Round, editor, \emph{Advances in Enzymology}, volume~2, pages
%  125--247. Academic Press, New York, 3rd edition, 1954.

%\bibitem[Keohane(1958)]{Keo:58}
%\newblock \emph{Power and Interdependence: World Politics in Transitions}.
%\newblock Little, Brown \& Co., Boston, 1958.

%\bibitem[Powers(1985)]{Pow:85}
%\newblock Is there a way out?
%\newblock \emph{Harpers}, pages 35--47, June 1985.

%\bibitem[Soukhanov(1992)]{Heritage:92}
%\newblock \emph{{The American Heritage. Dictionary of the American Language}}.
%\newblock Houghton Mifflin Company, 1992.

%\end{thebibliography}

                                                                         % in the appendices.
\end{document}